\documentstyle[prl,twocolumn,aps]{revtex}

\newcommand {\be}{\begin{equation}}
\newcommand {\ee}{\end{equation}}

\begin{document}

\draft

\title{A generating partition for the standard map}
\author{Freddy Christiansen$^{\clubsuit}$,
Antonio Politi$^{\clubsuit,\spadesuit}$
\quad\\
$\clubsuit$ {\it Istituto Nazionale di Ottica, I-50125 Firenze, Italy}\\
$\spadesuit$ INFN. {\it Sezione di Firenze}}

\twocolumn[
\date{\today}

\maketitle

\widetext

\vspace*{-1.0truecm}

\begin{abstract}
\begin{center}
\parbox{14cm}
{A procedure to obtain the symbolic dynamics for conservative dynamical
systems is introduced with reference to the standard map in a strongly
chaotic regime. The method extends an approach previously developed for
highly dissipative systems. It is based on the construction of a generating
partition from homoclinic tangencies and fibers of invariant manifolds. It
is found that some arbitrariness in the construction of the partition is
unavoidable.}
\end{center}
\end{abstract}
] \narrowtext

An effective representation of chaotic dynamics can be achieved by encoding
any trajectory as an infinite sequence of symbols. This enables
a fruitful mapping: orbits can formally be seen as microstates of some spin
chain (the symbols corresponding to spin values). Accordingly, a
thermodynamical formalism can be developed \cite{therm} to compute relevant
statistical averages such as Lyapunov exponents, dynamical entropies and
fractal dimensions \cite{eck}.

Various approaches have been introduced to encode a given trajectory
in phase space. One method relies on the assumption that the code assigned
to each periodic orbit remains unchanged when the dynamical system is
smoothly modified \cite{hansen2}. The key aspect of this is the
identification of some parameter value $k_h$ such that the resulting
dynamics is characterized by a complete horseshoe. The encoding of each
periodic orbit for the desired parameter value $k_0$ is obtained by smoothly
deforming the orbit from $k_h$ to $k_0$. Unfortunately, it has been
discovered that there exist periodic orbits which, followed along closed
paths in parameter space, are transformed into different orbits, thus showing
unavoidable ambiguities \cite{hansen}.
A second method is based on the simultaneous introduction of a
pseudo-dynamics along a formal time-axis and on the interpretation of the
true time variable $n$ as a spatial index \cite{wenzel}. The applicability of
this approach is limited to strongly dissipative models.

A last powerful method, which works whenever a horseshoe type mechanism is
present in the dynamics, is based on the direct construction of a generating
partition (GP) by connecting together the relevant (primary) homoclinic
tangencies (HT) to eventually split the phase space into disjoint atoms
\cite{kantz}. Such a strategy has been successfully applied to both
maps and flows \cite{giovan} and it appears to be of general validity,
although there is no rigorous proof that it is always applicable. However,
this method too has been implemented only in dissipative systems. In fact,
for the Hamiltonian case serious difficulties arise in connecting the primary
HTs to form continuous partition lines.

A complete encoding of the dynamics in a conservative system requires taking
into account stability islands as well as the chaotic component in which
they are embedded. The former problem can in principle be solved by encoding
the rotation angles with respect to suitable reference points. An approach
in this direction has been developed by Russberg \cite{russb} for the
piecewise linear standard map in a regime where the phase space is
essentially filled by islands. Here, we focus our attention on the
complementary problem of a correct description of the chaotic evolution.
To this aim we have studied the standard map for a large nonlinearity, such
that the stability islands cover a tiny portion of the phase space. In
particular, we describe a procedure to identify and connect primary HTs in
such a way that the resulting line represents the border of a generating
partition.

Let us first briefly recall the main ideas behind the method originally
proposed in Ref. \cite{kantz}. Because of the folding process associated
with a horseshoe, if fibers of the unstable ($W_u$) and stable ($W_s$)
manifolds intersect each other, they must do so twice except for points
of tangency. The trajectories stemming from any pair of intersections
approach each other
both in the past and in the future, as they belong to the same branch
of both $W_s$ and $W_u$. Since the same reasoning applies to
close pairs of nearly tangent
intersections as well, it follows that the only way to distinguish the
corresponding symbolic sequences is to set the border of the partition
either on the tangency point, or on some backward (forward) image of
it. As long as one limits the analysis to just one fiber, all choices are
equivalent. However, the partition of phase space into distinct atoms
requires taking all fibers simultaneously into account. As a consequence,
one is faced with the problem of identifying the ``primary'' tangencies as
those effectively used to construct the GP. In practice, one starts with an
Ansatz about the region which is expected to contain the primary tangencies
(typically the folding region of the horseshoe). Then, different tangencies
are connected by following a sort of trial and error approach.

The standard map represents a simple but general model for testing methods
to analyse Hamiltonian systems. We write the transformation $F$ as
\begin{eqnarray}
\label{map}
    x_{n+1} & = &  y_n \nonumber \\
    y_{n+1}   & = & -x_n+2y_n  - \alpha \cos(y_n ) \bmod{2 \pi}  \quad .
\end{eqnarray}
We have chosen the value $\alpha=6$ throughout the Letter.
The variables $x$ and $y$ have been introduced in place of the commonly
used $\theta = x$ and $\rho = y - x$, since the resulting representation
guarantees that horizontal lines are mapped onto vertical lines, thus making
the partition look more natural. Let us note that map (\ref{map}) is
invariant under the composition of a time reversal plus the exchange of
$x$ and $y$ variables, and under the transformation $(x,y) \rightarrow
(\pi-x,\pi-y)~(\bmod{2 \pi})$.

The map exhibits two folding regions situated approximately at the maximum
and minimum of the curve $F(x_0,t)=(t,-x_0+2t-\alpha \cos(t))$, i.e. at the
vertical lines defined by $x=\sin^{-1}(-2/\alpha)$, of which we specifically
choose the two lines $L_1:(x=3.481\dots)$ and $L_2:(x=5.943\dots)$. These
two lines will be the basis for the construction of an approximate
generating partition.

Since the phase space is a torus, there are no natural boundaries along both
$x$ and $y$ directions. One must, therefore, break the continuity by
introducing two sets of transversal lines separated by a
distance $2\pi$ horizontally and vertically, respectively. This can, for
instance, be done by using the vertical line $L_2$ and its horizontal
preimage $F^{-1}L_2$. As a result, the plane is partitioned into infinitely
many equivalent squares $S$. Any other pair of transversal lines is, in
principle, equivalent; the idea of using a folding line such as $L_2$
is inspired by an attempt to minimize the number of partition elements.
If the second folding line $L_1$ is also used, then $S$ is split into two
elements. The resulting partition is not sufficiently fine-grained to
account for the multiplicity of trajectories generated by map (\ref{map}).
In fact, the (pre)images of the two elements intersect different copies of
$S$. One is therefore led to split each of the two elements into as many
atoms as the number of copies of $S$ which are visited. This is automatically
obtained by using $F L_2$ as a further dividing line. As a result, one
obtains a partition which should be approximately generating (see Fig. 1).
In fact, a check done with periodic orbits of increasing period shows that
a large fraction of them is correctly discriminated by the above partition.
There are, however, a number of orbits described by the same code. This
problem is not at all unexpected, since the partition has been constructed
starting from rather arbitrary lines identified by just looking at one
application of the map; it is well known that a HT involves an infinity
of steps.

Before starting the discussion about the refinement of $L_1$ and $L_2$,
let us notice that the line $L_2$ can be transformed into $L_1$ by exploiting
the symmetry of map (\ref{map}). We can, therefore, limit
ourselves to the study of one folding line. Starting from the two hyperbolic
fixed points $(\pi/2,\pi/2)$ and
$(3\pi/2,3\pi/2)$ of map (\ref{map}), we have constructed their respective
unstable manifolds by computing the coefficients of suitable power-series
expansions and iterating the resulting fibers. HTs have then been located by
determining the curvature of the unstable manifold at forward iterates of
points on the unstable manifold in the vicinity of $L_1$ \cite{giovan}. In
fact, a HT, being just a folding point, is characterized by a diverging
curvature.

Such a procedure leads to a tentative set of primary HTs which are seen to
align approximately along $L_1$. In
analogy with dissipative maps, it appears natural to connect such points in
ascending order, according to their $y$-coordinate. Although the resulting
curve is somewhere relatively smooth, discontinuities are clearly visible.
In dissipative maps, this is not considered to be a serious problem. The fact
that the attractor does not fill the whole phase space hands one a large
degree of freedom in connecting HTs that are far apart, as long as they are
not separated by pieces of the attractor. This is no longer true in a
conservative map, where the entire phase space is
typically filled by a single ergodic component (with the exception of
stability islands which need to be considered separately).

In order to better clarify what happens around each discontinuity, let us
look closer at one example, namely the pair of jumps indicated by arrows in
Fig. 1 and depicted in Fig. 2. We realize that the jump is the consequence
of an avoided crossing between two lines of HTs. The
discontinuity is in fact caused by the intersection of what will become a
border of our generating partition with a forward or backward image of
itself. Such a phenomenon is clearly seen in Fig. 2 where forward and
backward images (dotted lines) of the ``primary'' tangencies (solid lines)
have been added (the region in Fig. 2b is the second iterate of that
depicted in Fig. 2a).

Therefore, one is faced with the question of which HTs should be used
to discriminate different trajectories. Some degree of arbitrariness is
apparent for tangencies which return back to the folding region.
In principle, discontinuities arising from the intersections of a dividing
line with forward and backward images of itself are present everywhere, but
the jumps appear to diminish with the respective number of iterates needed to
return to the folding region. We can therefore attack this problem starting
from the larger gaps.

In Fig. 2b it is seen that three distinct tangencies are identified on those
fibers of $W_u$ which are not too close to the jump. The first
and the last of such points are unambiguously classified as primary points,
whereas the middle one corresponds to the 2nd iterate of a tangency
classified as primary (in Fig. 2a). Upon shifting the fiber of reference
towards the critical region, the two lower HT's meet and eventually
disappear, preventing a continuation of the dividing line. This process
was already discovered in dissipative maps upon changing a control
parameter \cite {giovan2}. In particular, it was shown how it is
associated with the difficulty of providing a unique characterization of
the symbolic encoding of periodic orbits \cite{hansen}. In a conservative
system, like the standard map under investigation, the same problem occurs
for any parameter value, since moving with continuity across the fibers of
$W_u$ is like changing a parameter of the dynamics.

{}From the point $Q_2$, where two strands of HTs collide,
one would like to find a way to
connect the partition to the nearby sequence of HTs, thus bridging the gap
arising from the apparent avoided crossing. Let us focus our attention on
the closed region $U$ delimited by the line of HTs between $R_2$ and $P_2$
and by the fibers of stable and unstable manifold departing from $Q_2$.
With reference to Fig. 2, it is
seen that trajectories visiting $U$ can be discriminated against
companion orbits (lying on the opposite side of a dividing line) either
when they lie in $F^{-2}U$, or when they are in $U$ itself. We
conjecture that any curve $C$ lying in $U$ and connecting $Q_2$ with a point
$S$ on the strand of HTs between $P_2$ and $R_2$ is appropriate, provided
that $F^{-2}C$ also is used in $F^{-2}U$ in a self-consistent manner. Two of
the infinitely many possible choices for $C$ appear to be most natural:
$W_s$ and $W_u$ themselves. This same ambiguity arises for any point on
the dividing line which returns to the folding region. Thus, we have an
infinity of bubbles analogous to $U$. It is therefore convenient to adopt
everywhere the same choice. The line $D_1$ resulting from the application
of this procedure to the larger gaps is plotted in Fig. 3, where fibers of
the unstable manifold have been used.

A generating partition can be constructed by using $D_1$ and its symmetric
equivalent $D_2$ analogously to the construction of the preliminary
partition from the lines $L_1$ and $L_2$. This finally results in a seven
letter alphabet as shown in Fig. 3. We have
tested the partition on all periodic orbits up to length 9 ($\approx 30,000$
orbits)
and found that the symbol sequences were unique except for a period-6 orbit
and four period-8 orbits around a stable period 2 region, sharing that of
the mother orbit. A correct encoding of such orbits
requires an {\it ad hoc} treatment of the corresponding stability island
\cite{russb}.

{}From the existence of 7 different period-2 orbits, it turns out that at least
5 symbols are needed for a correct encoding of the dynamics.
One might try to combine some of the atoms of the 7 letter alphabet of Fig. 3
into larger elements. However, from the study of all possible combinations of
atoms it is verified that only the triangular region appearing at small
$y$-values can be assimilated to another region without loss of information.
It is therefore very likely that 6 represents the minimum number of symbols.

A further check of the correctness of the partition has been performed by
estimating the Kolmogorov-Sinai entropy from the block entropies
\be
   H_k = - \sum_i p_i(k) \log p_i(k)
\label{blo}
\ee
where the sum is taken over all sequences of length $k$ and comparing them
with the numerical estimate of the maximum Lyapunov exponent
($\lambda \simeq 1.1365\dots$). From the $H_k$ values computed for $k \le 11$
and
reported in Table 1, it appears that they are converging (in the accessible
$k$ range) slower than exponentially and faster than algebraically. By taking
this into account, the
extrapolated value of $H_\infty$ is in good agreement with $\lambda$, as
required from the Pesin relation. We can therefore conclude that the
partition constructed in this Letter is a good generating partition
which can be used for encoding trajectories and thus improving the
understanding of conservative non-hyperbolic systems. In particular,
it is now possible to attack the problem of constructing a pruning front
\cite{front}.

\begin{table}
\begin{tabular}{|r|l|}
\multicolumn{1}{|c|}{$k$} & \multicolumn{1}{|c|}{$H_k$}  \\ \hline
 1 & 1.77292 \\
 2 & 1.69554 \\
 3 & 1.59844 \\
 4 & 1.52601 \\
 5 & 1.47195 \\
 6 & 1.43028 \\
 7 & 1.39680 \\
 8 & 1.36969 \\
 9 & 1.3475  \\
10 & 1.326 \\
11 & 1.312 \\
\end{tabular}
\caption[tabone]{Block entropies $H_k$ defined in Eq. (\ref{blo})
for lengths $k\le 11$.}
\label{tab1}
\end{table}

\begin{figure}
\caption[figone]{Approximate generating partition. The primary region of
phase space is obtained by using the vertical line $L_2$ (solid) and its
preimage $F^{-1}L_2$ (dotted). This region is then partitioned by $L_1$
(solid) and the image of $L_2$ (dashed). Dots denote homoclinic tangencies
classified as primary according to their vicinity to $L_1$. The arrows
point to the regions reported in Fig. 2.}
\label{fig1}
\end{figure}

\begin{figure}
\caption[figthree]{Enlarged picture of the avoided crossings at the two
arrows in Fig. 1; Fig. 2b is the second forward image of Fig. 2a. The
solid-dotted lines refer to HTs: solid parts denote tangencies unambiguously
identified as primary; the points along $P_0R_0$ and $R_2P_2$
may be taken as primary in either region (but not both). $Q_0$ and its
second image $Q_2$ are identified as the points where two sequences of
HTs meet and collapse. The stable and unstable manifolds (dashed lines)
departing from $Q_0$ ($Q_2$) intersect the strand of HTs in $P_0$ ($P_2$)
and $R_0$ ($R_2$), respectively. A branch of the unstable manifold
containing three tangencies (triangles) is also shown in (b).}
\label{fig3}
\end{figure}

\begin{figure}
\caption[figfour]{The generating partition as constructed from primary HTs
and suitable pieces of unstable manifolds. In analogy to Fig. 1 we have used
the dividing line $D_2$ (solid) and its preimage (dotted) to define
the primary region of phase space. This is then partitioned by $D_1$ (solid)
and the image of $D_2$ (dashed).}
\label{fig4}
\end{figure}

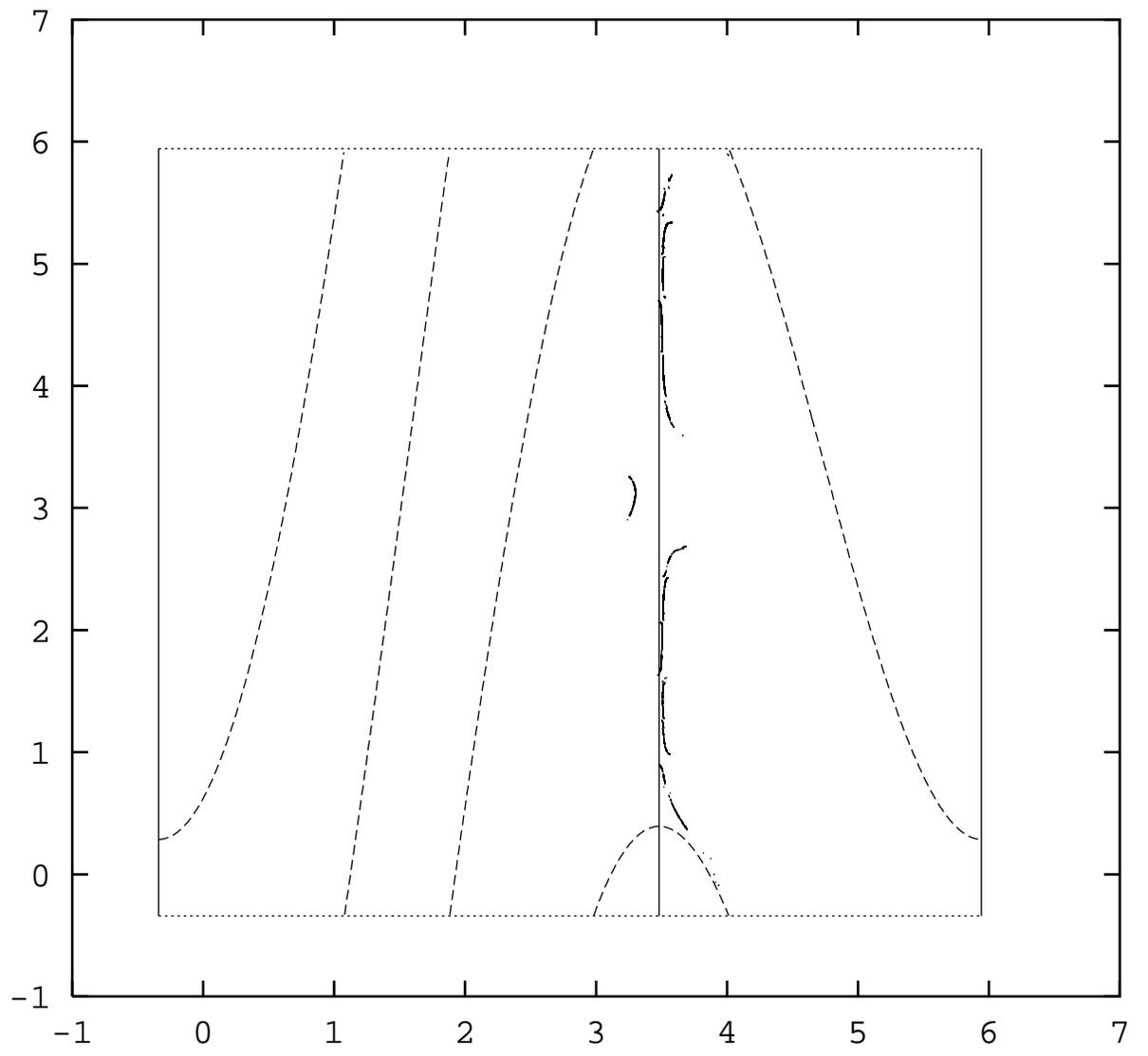
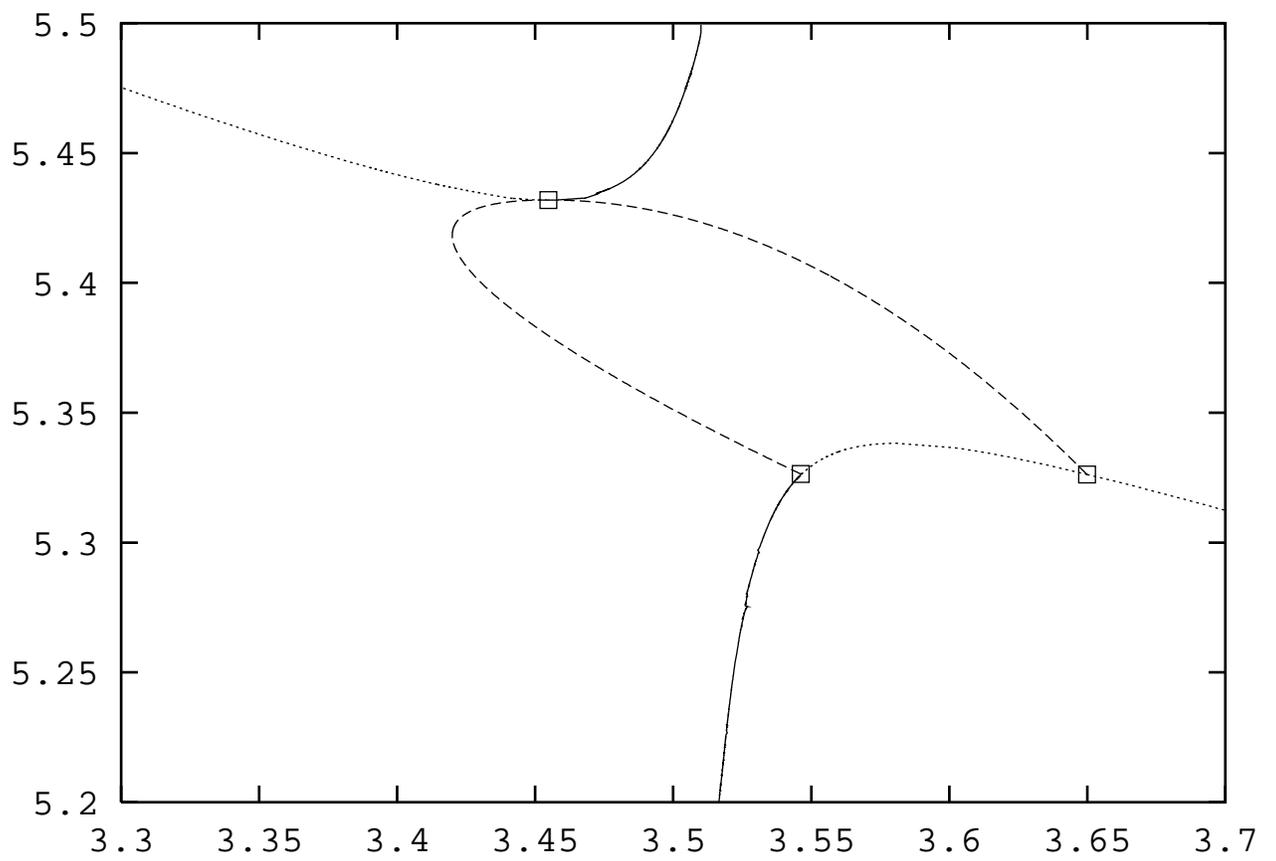
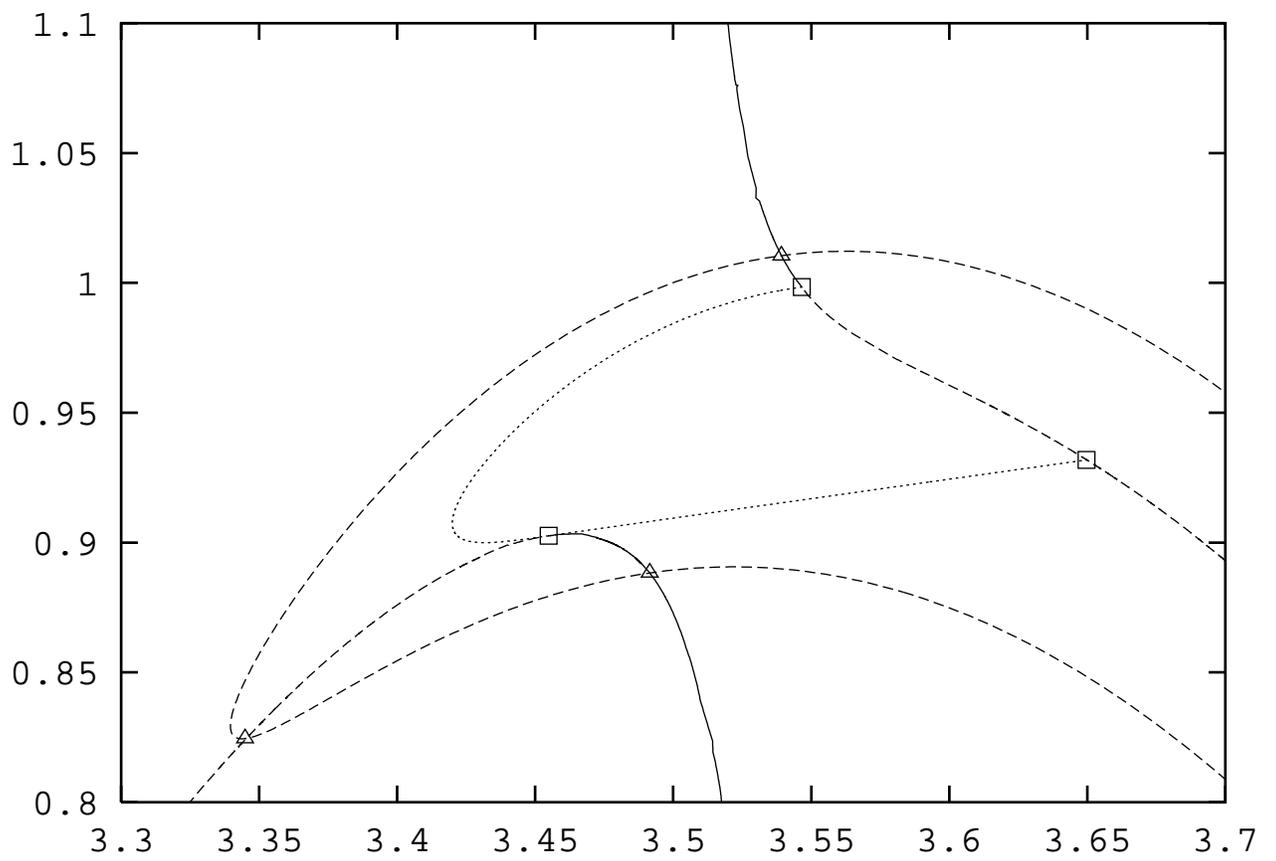
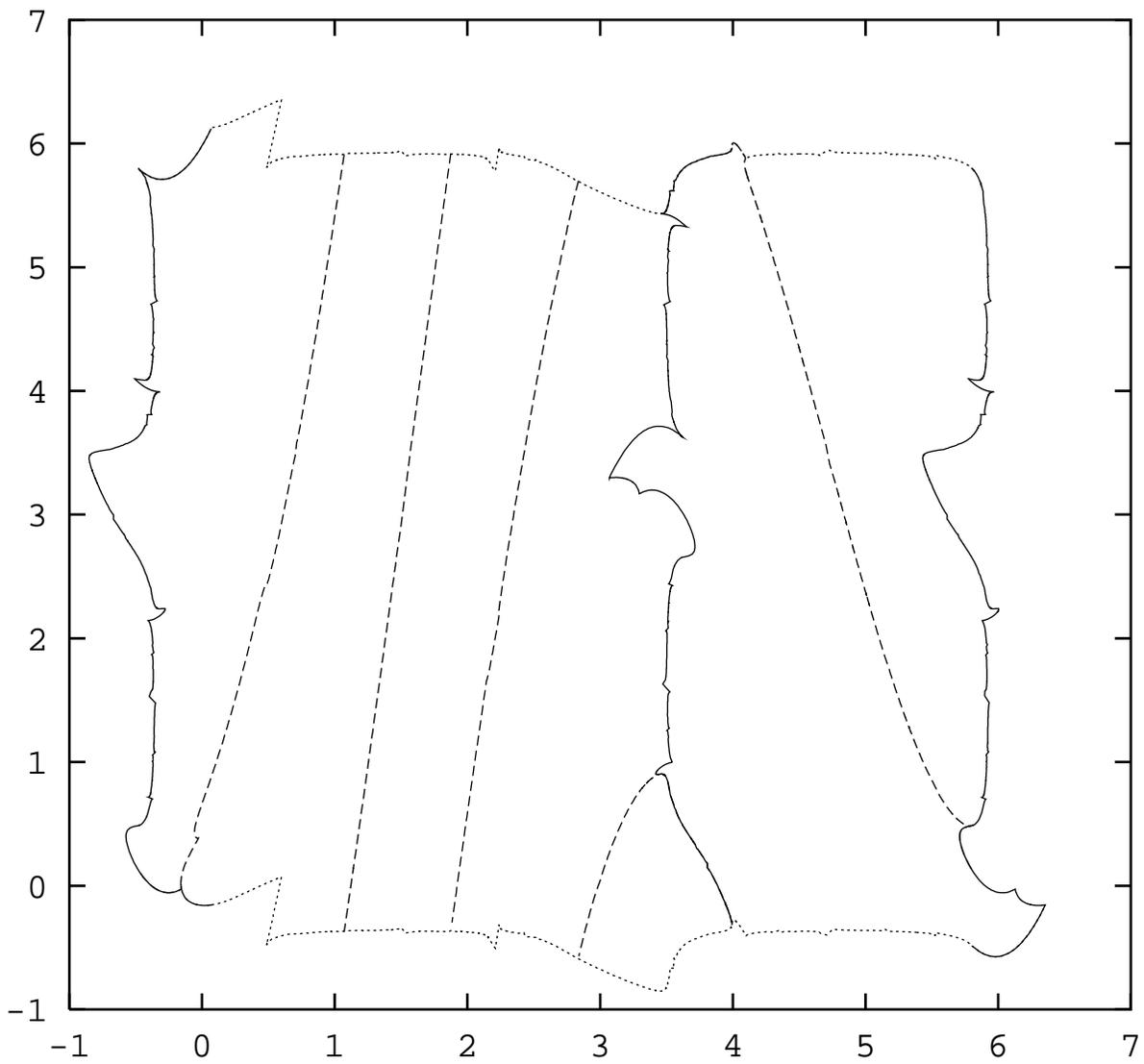
\end{document}